\documentclass[pra,aps,twocolumn,nopacs,superscriptaddress,nofootinbib,longbibliography]{revtex4-1}

\usepackage{amsmath}  \usepackage{amssymb}  \usepackage{amsfonts}  \usepackage{bm}  \usepackage{bbm}   \usepackage{bbold}  \usepackage{braket}  \usepackage{color}  \usepackage{comment}  \usepackage{dcolumn}  \usepackage{enumerate}  \usepackage{epsfig}  \usepackage{gensymb}  \usepackage{graphicx}  \usepackage{indentfirst}  \usepackage{lmodern}  \usepackage{mathrsfs}  \usepackage{mathtools}  \usepackage{psfrag}  \usepackage{pst-all}  \usepackage{soul}  \usepackage{xcolor}
\usepackage{float}  \usepackage[colorlinks,linkcolor=blue,citecolor=blue,urlcolor=blue,hyperindex,driverfallback=dvipdfm]{hyperref}  \usepackage[T1]{fontenc} 
\def\ii{{\rm i}}  \def\ee{{\rm e}}
  \def\kB{{k_{\rm B}}}
  \def\Imm{{\rm Im}}
      
  \def\eb{{\bf e}}  \def\Fb{{\bf F}}  
      
      \def\kb{{\bf k}}

    \def\vb{{\bf v}}
    \def\zz{\hat{\bf z}}
    \def\eh{\hat{\bf e}}
    
    \def\kperb{{\bf k}_\perp}
\def\abs{\rm{abs}}  \def\scat{\rm{scat}}
\def\pabs{\mathcal{P}_{\abs}}  \def\pscat{\mathcal{P}_{\scat}}

\begin{document}

\title{Thermal vacuum friction of objects with different dimensionality}

\author{Gor~Chalyan}
\affiliation{ICFO-Institut de Ciencies Fotoniques, The Barcelona Institute of
Science and Technology, 08860 Castelldefels (Barcelona), Spain}

\author{F.~Javier~Garc\'{\i}a~de~Abajo}
\email{javier.garciadeabajo@nanophotonics.es}
\affiliation{ICFO-Institut de Ciencies Fotoniques, The Barcelona Institute of
Science and Technology, 08860 Castelldefels (Barcelona), Spain}
\affiliation{ICREA-Instituci\'o Catalana de Recerca i Estudis Avan\c{c}ats,
Passeig Llu\'{\i}s Companys 23, 08010 Barcelona, Spain}

\begin{abstract}
Radiative forces acting on neutral bodies moving through a thermal bath represent a unique manifestation of the interplay between relativistic kinematics and thermal fluctuations. Vacuum friction is commonly formulated using the fluctuation--dissipation theorem or related statistical approaches, but such treatments can obscure the elementary momentum-transfer processes, especially in relativistic regimes. Here, we develop a kinetic momentum-transfer framework based on relativistic kinematics in which the radiative force and pressure are obtained by summing individual scattering and absorption events. This approach offers a transparent physical picture while ensuring a self-consistent treatment of Doppler shifts and relativistic transformations. We apply the method to three representative geometries: an isotropic dipolar particle, a thin resonant plate moving normal to its surface, and a thin resonant plate moving parallel to its surface. In the nonrelativistic limit, we derive explicit radiative drag coefficients, providing compact expressions for predicting vacuum friction in moving structures.
\end{abstract}
\maketitle
\date{\today}

\section{Introduction}\label{intro}

The interaction between moving objects and thermal radiation has long been a subject of fundamental interest in physics, lying at the intersection of thermodynamics, electrodynamics, and special relativity \cite{P1908,O1963,M1972_2}. An object moving through a thermal photon bath experiences a radiative drag force, a phenomenon commonly referred to as vacuum friction, arising from the asymmetric exchange of momentum with the surrounding thermal field. This net force, or radiative pressure, is governed by the absorption, reflection, and scattering of thermal photons by the moving object.

Historically, the study of radiative friction dates back to the seminal work of Einstein and Hopf \cite{EH1910}, who showed that an atom immersed in a blackbody radiation bath experiences a drag force proportional to its velocity. Subsequent theoretical work has extended this concept to a broad range of configurations and regimes. Microscopic damping has been associated with electronic and vibrational excitations \cite{GEM91,P91_2}. Modern interest in noncontact friction has been strongly motivated by experimental observations, including residual drag on atomic-force-microscope tips in vacuum \cite{G02}. This subject has stimulated extensive studies of Casimir dissipative drag, which arises due to electromagnetic fluctuations in an equilibrium reference frame with respect to which a body is moving \cite{L1989,GK97,GK98}.

The problem of noncontact friction has been investigated for bodies in linear motion both in vacuum \cite{MPP03,MF04,MPP04,SM22} and near substrates \cite{VP03_2,VP07,WDT16}, where evanescent fields and surface excitations play a central role. Early studies of these effects, particularly for parallel plates, introduced the notion of the electromagnetic vacuum as a viscous medium and derived general friction coefficients for resonant structures \cite{M95}. In near-field configurations, friction is often dominated by evanescent fields and is therefore highly sensitive to the separation and material properties of the bodies in relative motion \cite{AE1986,TW97}.

Much of the recent literature addresses vacuum and thermal friction using fluctuational electrodynamics \cite{PH13}, often in covariant form to ensure consistency in relativistic regimes. In particular, the covariant formulation of Ref.~\cite{PH13} describes the force on a moving polarizable particle in terms of field and material fluctuations connected through the fluctuation--dissipation theorem. The material response is treated in the particle rest frame, while the thermal radiation is specified in the bath frame, with the corresponding quantities connected through Lorentz transformations. Alternative formulations have addressed the microscopic origin and nonequilibrium thermodynamics of quantum friction \cite{FFL19,RIH20,RIB22,PIF26}, while its existence and physical interpretation have continued to motivate theoretical discussion \cite{MHB16,MPK25}. Quantum friction has also been investigated in two-dimensional materials \cite{FKD18}, nanoscale liquid flows \cite{KBB22}, and other quantum-field-theoretical settings \cite{HKY25}.

The concept of radiative drag has also been extended to rotational degrees of freedom \cite{paper157,paper166,BL15,IOR19}, including enhancement by surface photon tunneling \cite{XJL21}. Recent experiments have further provided evidence of noncontact Casimir friction \cite{XJS24}. More generally, vacuum friction and related dissipative forces remain the subject of active theoretical and experimental investigation \cite{PL09,P10,L10,P10_2,VP11}. In a related context, heated planar bilayers composed of contrasting reflecting and absorbing materials have been argued to produce a net thermal force \cite{paper442}.

In this article, we investigate the net radiative force and pressure acting on thin resonant plates and dipolar particles as they move with a uniform velocity vector $\vb$ through a thermal bath in vacuum. We employ a kinetic momentum-transfer approach and evaluate the force and pressure directly from the momentum transferred between absorbed, emitted, or scattered thermal photons and the moving object. As in covariant fluctuation-electrodynamics formulations \cite{PH13}, the material response is evaluated in the object rest frame, while photon frequencies and momenta are connected to those in the lab frame through Lorentz transformations. Our approach differs in that the force is constructed directly from individual photon-transfer events, allowing the absorption, emission, and scattering contributions to be identified separately. This derivation provides an intuitive physical picture of the drag mechanism, ensures a self-consistent treatment of relativistic effects, and identifies the corresponding equilibrium conditions.

The paper is organized as follows. In Sec.~\ref{kinematics}, we present a general kinetic momentum-transfer theory of thermal friction based on momentum-transfer summations and a general form of the body--radiation interaction probabilities. In Sec.~\ref{geometries}, we apply this theory to three representative systems: an isotropic dipolar particle (Sec.~\ref{particle}), a plate moving normal to its surface (Sec.~\ref{vertical}), and a plate moving parallel to its surface (Sec.~\ref{horizontal}). The low-velocity limit and the corresponding drag coefficients are discussed in Sec.~\ref{lowv}. We summarize our main results in Sec.~\ref{conclusions}.

\section{Kinetic Momentum-Transfer Theory of Thermal Friction}\label{kinematics}

We describe the thermal friction experienced by a structure moving through vacuum at temperature $T_0$ by considering individual photon scattering and absorption events, summed over the thermal occupation numbers of the relevant photon states. The structure moves with a uniform velocity vector $\vb$ relative to the lab frame, defined as that in which radiation photons follow a Bose--Einstein distribution $n_0(\omega)=1/(\ee^{\hbar\omega/\kB T_0}-1)=1/(\ee^{2\pi\omega/\theta_0}-1)$ at temperature $T_0$, where $\theta_0=2\pi\kB T_0/\hbar$ is the thermal frequency. The material is considered to sustain bosonic excitations characterized by a distribution $n_1(\omega)=1/(\ee^{\hbar\omega/\kB T_1}-1)$ at temperature $T_1$. We take $\vb\parallel\zz$ without loss of generality.

\begin{figure*}[t]
\centering
\includegraphics[width=1.0\textwidth]{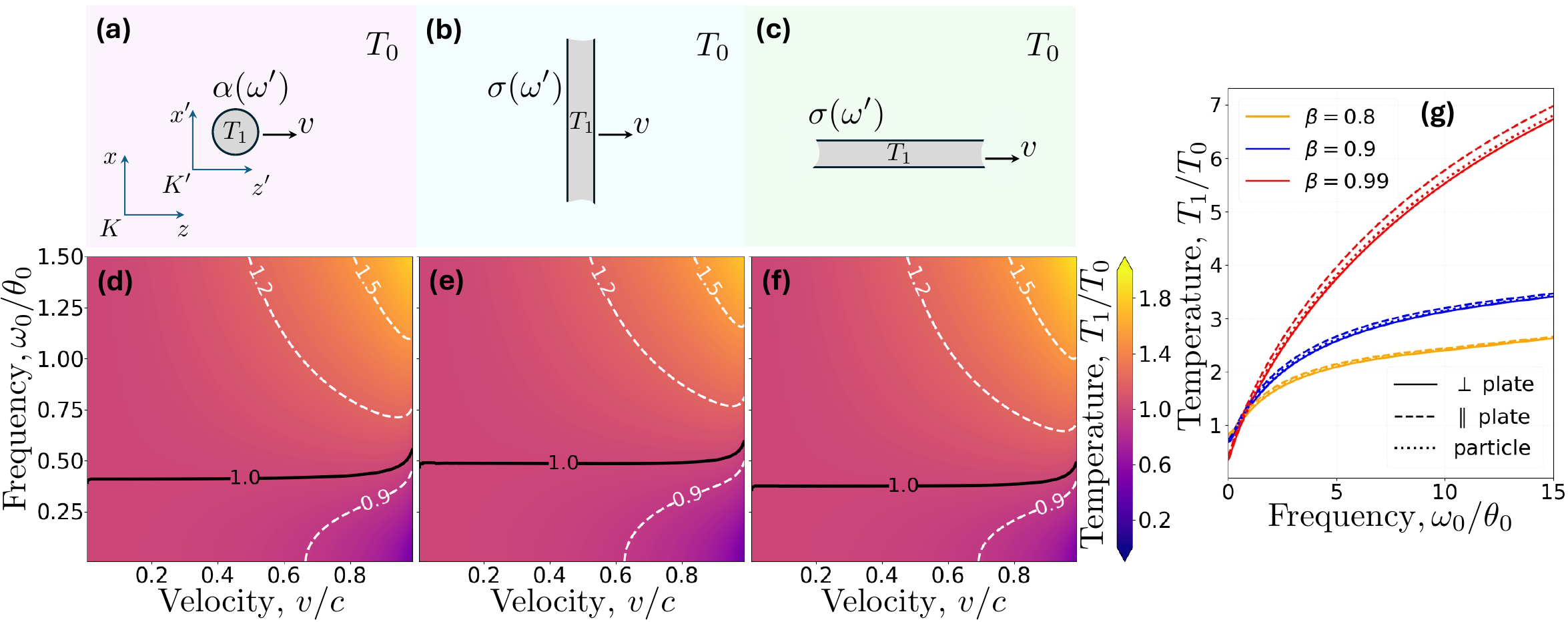}
\caption{Equilibrium temperature of a small particle and thin plates moving through a thermal radiation bath. We consider a small isotropic particle (a) and an extended plate moving either normal (b) or parallel (c) to its surface. The material response is assumed to consist of a narrow resonance at frequency $\omega_0$ in the rest frame. For the plates, the thickness is taken to be sufficiently small that scattering contributions to the resonance linewidth can be neglected. In panels (d)--(f), we show the ratio of the equilibrium material temperature $T_1$ to the vacuum temperature $T_0$ as a function of the normalized velocity $\beta=v/c$ and resonance frequency $\omega_0/\theta_0$, where $\theta_0=2\pi\kB T_0/\hbar$ is the thermal frequency, for the three configurations (a)--(c), respectively. The plate equilibrium conditions are evaluated for $\eta=0.4$. In panel (g), we show $T_1/T_0$ as a function of resonance frequency for the three types of moving objects and for different values of $\beta$, as indicated in the legend.}\label{Fig1}
\end{figure*}

Primed and unprimed symbols are used to denote quantities in the rest and lab frames [$K'$ and $K$, respectively; see Fig.~\ref{Fig1}(a)], while relativistic kinematics is used to transform photon wave vectors $\kb$ and frequencies $\omega=ck$ between the two frames. In particular, the wave vector $\kb=\kperb+k_z\zz$, where $\kperb=(k_x,k_y)$ is the component perpendicular to the direction of motion, transforms as \cite{J99}
\begin{subequations}\label{transform}
\begin{align}\label{kpk}
&\kperb'=\kperb,
\\\label{kpzk}
&k'_z=\gamma k\big(\mu-\beta\big),
\\\label{wpw}
&\omega'=\gamma\omega\big(1-\beta\mu\big),
\end{align}
where $\beta=v/c$ is the dimensionless velocity, $\gamma=1/\sqrt{1-\beta^2}$ is the Lorentz factor, and we define $\mu=\cos\theta$ with $\theta$ denoting the angle between $\kb$ and $\zz$. In what follows, we use the notation $\Omega=(\theta,\varphi)$ for the polar and azimuthal angles of $\kb$. Then, Eqs.~(\ref{kpk})-(\ref{wpw}) imply the transformations
\begin{align}\label{mupmu}
\mu'=(\mu-\beta)/(1-\beta\mu)
\end{align}
and $\varphi'=\varphi$. 
\end{subequations}

For the elastic contribution to friction, we describe each photon in the rest frame of the structure and calculate its scattering by the geometry under consideration, as detailed in Sec.~\ref{geometries}. The interaction is characterized by an angle-dependent scattering probability $\pscat^{\nu\nu_s}(\kb',\Omega'_s)$ for a photon with wave vector $\kb'$ and polarization $\nu$ in the rest frame to be redirected into the direction $\Omega'_s$ with polarization $\nu_s$ and frequency $\omega'_s=\omega'$ (because the scattering is elastic). We consider two orthogonal polarization states defined with respect to the velocity direction: p polarization has its electric field in the plane spanned by the velocity and the photon wave vector, whereas s polarization has its electric field perpendicular to this plane. This polarization labeling is preserved under the Lorentz boost between the lab and object rest frames (i.e., $\nu'=\nu$). A simple kinematic analysis based on Eqs.~(\ref{wpw}) and (\ref{mupmu}) shows that the lab-frame frequency of the scattered photon is determined by the incident frequency and the direction of scattering as
\begin{align}\label{wsw}
\omega_s=\gamma\omega'\,(1+\beta\mu'_s)=\gamma^2\omega\,(1-\beta\mu)(1+\beta\mu'_s),
\end{align}
where $\mu'_s=\cos\theta'_s$ in the rest frame. In Sec.~\ref{geometries}, the scattering probability is expressed in terms of the scattering cross section for small isotropic particles or the reflectivity for extended planar plates.

For the absorption contribution, we introduce the probability $\pabs^\nu(\kb')$ defined as the fraction of incident photons with wave vector $\kb'$ in the rest frame that excite internal degrees of freedom of the structure. This process removes the incident thermal photon and, therefore, transfers both momentum and energy. In Sec.~\ref{geometries}, we determine $\pabs^\nu(\kb')$ from the absorption cross section for small particles or the absorption for extended plates.

Putting these contributions together and summing over the wave vector $\kb$ and polarization $\nu$ of the photon distribution, we obtain the following expression for the radiative force:
\begin{align}\label{force}
&\Fb=\sum_{\kb,\nu} \frac{\hbar}{\tau_{\kb}}\, \Big\{\kb\big[n_0(\omega)-n_1(\omega')\big]\, \pabs^\nu(\kb')
\\\nonumber
&+\sum_{\nu_s}\!\int\!d^2\Omega'_s\big(\kb - \kb_s\big)\,n_0(\omega)\big[n_0(\omega_s)\!+\!1\big]\, 
\pscat^{\nu\nu_s}(\kb',\Omega'_s)\Big\}.
\end{align}
Here, the first and second terms describe absorption plus emission and scattering, respectively. Each term is multiplied by the corresponding momentum change in the lab frame [$\pm\hbar\kb$ and $\hbar(\kb-\kb_s)$, respectively] as well as the photon collision rate $\tau_{\kb}^{-1}$ (see below). For scattering, we also sum over final rest-frame directions $\Omega'_s$ and polarizations $\nu_s$, weighting each contribution by the occupation number $n_0(\omega)$ at the incident lab-frame frequency $\omega$ and by the stimulated-emission factor $[n_0(\omega_s)+1]$ at the scattered lab-frame frequency $\omega_s$ [Eq.~(\ref{wsw})]. The latter factor accounts for the addition of one photon to the mode $\{\kb_s,\nu_s\}$. For absorption, which transfers momentum $\hbar\kb$, the weighting factor $[n_1(\omega')+1]\,n_0(\omega)$ accounts for the removal of one photon from the radiation bath and the creation of a material excitation at the rest-frame frequency $\omega'$. Conversely, photon emission, which transfers momentum $-\hbar\kb$, is associated with the annihilation of a material excitation and is weighted by $n_1(\omega')\,[n_0(\omega)+1]$. By virtue of reciprocity, the probability $\pabs^{\nu}(\kb')$ is the same for absorption and emission. The net thermal factor in the first term of Eq.~(\ref{force}) is therefore $n_0(\omega)-n_1(\omega')$. Finally, we note that probabilities are evaluated in the rest frame, while frequencies, directions, and rates are transformed to the lab frame.

It should be noted that Eq.~(\ref{force}) can be regarded as a kinetic description in terms of individual photon--material interaction events. The momentum transferred in each event is weighted by the corresponding interaction probability and by the occupation factors of the initial and final states. In particular, detailed balance between absorption and emission gives rise to the factor $n_0(\omega)-n_1(\omega')$, while scattering is weighted by the occupation of the incident photon state and the bosonic factor associated with the final state. The present formulation applies when the interaction with the radiation field can be described in terms of the absorption and elastic-scattering probabilities introduced above.

To evaluate the sum over $\kb$, we introduce a quantization box of volume $L^3$ and take the limit $L\to\infty$. We also normalize the momentum transfer by the time $\tau_{\kb}$ between consecutive photon--material encounters for each mode with wave vector $\kb$ in the lab frame. Equivalently, $\tau_\kb$ is the time required for the moving object to accumulate a phase shift corresponding to the quantization length $L$. The phase of the photon plane wave at an instantaneous position of the object $z(t)=vt$ is $\phi(t)=\omega t-k z(t)=ckt(1-\beta\mu)$, and thus, we impose the condition $\phi(t+\tau_\kb)-\phi(t)=kL$, which readily leads to $\tau_{\kb}=(L/c)/(1-\beta\mu)$.

We are also interested in the material temperature $T_1$ under dynamical equilibrium conditions, for which the power $\dot{Q}$ deposited into the internal particle energy vanishes. Following an analysis analogous to that leading to Eq.~\eqref{force}, we obtain
\begin{align}\label{equilibrium}
\dot{Q}=\sum_{\kb,\nu} \frac{\hbar\omega'}{\tau_{\kb}}\, \big[n_0(\omega) - n_1(\omega')\big]\, \pabs^\nu(\kb')
\end{align}
for the internal power expressed in the lab frame. Here, $T_1$ is defined in the rest frame of the moving object. We assume that internal relaxation is sufficiently fast compared with the radiative heating and mechanical time scales, so that the internal degrees of freedom remain in thermal equilibrium and can be described by a Bose--Einstein distribution at temperature $T_1$. Under this assumption, the condition $\dot{Q}=0$ determines the equilibrium temperature of the object. If internal thermalization is restricted to the resonant degree of freedom considered in our model below, $T_1$ should instead be understood as an effective temperature associated with that mode.

Notice that $T_1$ is determined from the energy balance rather than from a relativistic transformation of the bath temperature $T_0$. The latter is defined in the lab frame, where the radiation distribution is isotropic, while the moving object experiences direction-dependent Doppler shifts. This generally leads to an equilibrium value $T_1\neq T_0$, as we discuss below.

\section{Application to selected geometries}\label{geometries}

We apply the formalism presented in Sec.~\ref{kinematics} to three types of moving structures, as sketched in the upper part of Fig.~\ref{Fig1}: (a)~an isotropic particle, (b)~an extended homogeneous plate moving normal to its surface, and (c)~a plate moving along an in-plane direction. Each structure is analyzed in a separate subsection. The treatment of polarization under scattering depends on the geometry. In particular, the p and s states defined above with respect to the velocity direction are not, in general, preserved upon scattering. The polarization sums entering Eqs.~(\ref{force}) and (\ref{equilibrium}) are evaluated below in the basis most convenient for each geometry. We compute the force and energy transfer from Eqs.~(\ref{force}) and (\ref{equilibrium}), using the prescription $\sum_\kb\to (L/2\pi)^3\int d^3\kb$ for the sums over wave vectors and implicitly applying the transformations in Eqs.~(\ref{transform}) throughout.
\subsection{Small isotropic particle}\label{particle}

We consider a small isotropic particle whose size is much smaller than the wavelengths of the photons under consideration, so that its optical response can be described by a frequency-dependent polarizability $\alpha(\omega)$. In the rest frame, absorption and scattering are characterized by their respective cross sections \cite{NH06}
\begin{subequations}\label{cross_section}
\begin{align}
&\sigma'_{\abs}(\omega') = 4\pi k'\left[\Imm\{\alpha(\omega')\}- \frac{2k'^3}{3}|\alpha(\omega')|^2\right], \\
&\sigma'_{\scat}(\omega') = \frac{8\pi k'^4}{3}|\alpha(\omega')|^2,
\end{align}
\end{subequations}
which are independent of polarization. The per-photon absorption probability is then given by the ratio  $\pabs^\nu(\kb')=L^{-2}\sigma'_{\abs}(\omega')$ between the absorption cross section and the transverse area of the quantization box.

The scattering probability, however, requires a more detailed analysis. In the rest frame, the angular distribution of scattered photons follows the radiation pattern of a dipole oriented along the incident polarization vector $\eh'_\nu$. In general, scattering into a different propagation direction can mix the p and s polarization states defined with respect to the velocity direction. Because the momentum transferred to the particle is independent of the polarization of the scattered photon, we sum over the final polarization and define $\mathcal{P}_{\rm scat}^{\nu}(\kb’,\Omega’_s)
\equiv\sum_{\nu_s}\mathcal{P}_{\rm scat}^{\nu\nu_s}(\kb’,\Omega’_s)\sigma’_{\rm scat}(\omega’),g_\nu(\Omega’,\Omega’_s)$. The resulting polarization-summed angular distribution is that of an electric dipole oriented along $\eb’_\nu$, namely $g_\nu(\Omega’,\Omega’_s)\propto1-|\eb’_\nu\cdot\hat{\kb}’_s|^2$. Since the system has axial symmetry about the $z$ axis, we consider the azimuthal integral of $g_\nu(\Omega’,\Omega’_s)$. Enforcing the normalization condition $\int d^2\Omega’_s\,g_\nu(\Omega’,\Omega’_s)=1$, we obtain
\begin{align}\nonumber
\sum_\nu\int_0^{2\pi}d\varphi'_s\;g_\nu(\Omega',\Omega'_s) =\frac{3}{8}\big(3-\mu'^2-\mu'^2_s+3\mu'^2\mu'^2_s\big),
\end{align}
which is independent of $\varphi'$.

Combining the above ingredients, Eqs.~(\ref{force}) and (\ref{equilibrium}) reduce to
\begin{widetext}
\begin{subequations}\label{FPparticle}
\begin{align}\nonumber
&F=\!\frac{\hbar}{2\pi^2c^3} \int_0^\infty\!\!\!\omega^3d\omega\!\int_{-1}^1\!\!\!d\mu\, (1\!-\!\beta\mu)
\,\bigg\{\mu\big[n_0(\omega)\!-\!n_1(\omega')\big] \sigma'_{\abs}(\omega')
\\\label{Fparticle}
&\quad\quad\quad\quad\quad\quad\quad\quad\quad +\frac{3}{16}\int_{-1}^1 \!\!d\mu'_s\, \frac{\mu\!-\!\mu_s}{1\!-\!\beta\mu_s}\, \big(3-\mu'^2-\mu'^2_s+3\mu'^2\mu'^2_s\big)\,
n_0(\omega)\big[n_0(\omega_s)\!+\!1\big]
\sigma'_{\scat}(\omega') \bigg\},
\\\label{Pparticle}
&\dot{Q}=\!\frac{\hbar\gamma}{2\pi^2c^2} \int_0^\infty\!\!\!\omega^3d\omega\!\!\int_{-1}^1\!\!d\mu\, (1\!-\!\beta\mu)^2 \big[n_0(\omega)\!-\!n_1(\omega')\big] \sigma'_{\abs}(\omega'),
\end{align}
\end{subequations}
\end{widetext}
where $\omega'$, $\mu'$, and $\omega_s$ are defined in Eqs.~(\ref{wpw}), (\ref{mupmu}), and (\ref{wsw}), respectively, while $\mu_s=\beta+\mu'_s\omega'/\gamma\omega_s$ is the cosine of the scattering angle in the lab frame.

For illustration, we consider a particle with a rest-frame response dominated by a single resonance mode of frequency $\omega'=\omega_0$. Its polarizability is modeled as a Lorentzian function of the form \cite{VST96}
\begin{align}\label{alpha}
\alpha(\omega')=\frac{3\,c^3\kappa_r}{2\,\omega_0^2}
\frac{1}{\omega_0^2-\omega'\big[\omega'+\ii\kappa(\omega'/\omega_0)^2\big]},
\end{align}
where $\kappa$ denotes the total mode decay rate and $\kappa_r$ is the radiative contribution to $\kappa$. The expression in Eq.~(\ref{alpha}) is constructed so that the optical theorem gives ${\rm Im}\{-1/\alpha(\omega_0)\}=(\kappa/\kappa_r)\,2\omega_0^3/3c^3$ at resonance ($\omega'=\omega_0$), corresponding to a perfect scatterer when $\kappa_r=\kappa$. In what follows, we assume a narrow linewidth $\kappa\ll\omega_0$, such that we can transform
\begin{align}\label{deltafunction}
\frac{\kappa/2}{(\omega'-\omega_0)^2+(\kappa/2)^2}\to\pi\delta(\omega'-\omega_0),
\end{align}
in the cross sections [Eqs.~(\ref{cross_section})]. This allows us to evaluate the remaining factors in the $\omega$ integrands of Eqs.~(\ref{FPparticle}) at the resonant rest- and lab-frame frequencies $\omega'\approx\omega_0$ and $\omega\approx\omega_0/\gamma(1-\beta\mu)$, respectively. Then, the force becomes
\begin{align}\label{FFF}
F=F_{\rm abs}+F_{\rm scat},
\end{align}
where
\begin{subequations}\label{Fparticlefinal}
\begin{align}\label{Fabs}
&F_{\rm abs}\approx\frac{3\,\hbar\omega_0\,\kappa_r}{2c}\Big(1-\frac{\kappa_r}{\kappa}\Big) G_{\rm abs},
\\\label{Fscat}
&F_{\rm scat}\approx\frac{3\,\hbar\omega_0\,\kappa_r}{2c}\,\frac{\kappa_r}{\kappa} G_{\rm scat}
\end{align}
\end{subequations}
are the absorption and scattering components, written in terms of the dimensionless functions
\begin{subequations}\label{GGG}
\begin{align}\label{Gabs}
&G_{\rm abs}=\frac{1}{\gamma^4} \int_{-1}^1\!\frac{\mu\,d\mu}{(1-\beta\mu)^3}\, \big[n_0(\omega)-n_1(\omega_0)\big],
\\\label{Gscat}
&G_{\rm scat}=\frac{3}{16\gamma^4} \int_{-1}^1\!\frac{d\mu}{(1-\beta\mu)^3} \int_{-1}^1 \!d\mu'_s\, \frac{\mu-\mu_s}{1-\beta\mu_s}
\\\nonumber
&\quad\quad\quad\times
\big(3-\mu'^2-\mu'^2_s+3\mu'^2\mu'^2_s\big)\, n_0(\omega)\, \big[n_0(\omega_s)+1\big].
\end{align}
\end{subequations}
In Appendix~\ref{GisG}, we show that Eqs.~(\ref{GGG}) can be written in a simpler form as
\begin{subequations}\label{GGGeq}
\begin{align}\label{Gabseq}
&G_{\rm abs}=\int_{-1}^1\!d\mu'\,(\beta+\mu')\, \big[n_0(\omega)-n_1(\omega_0)\big],
\\\label{Gscateq}
&G_{\rm scat}=\int_{-1}^1\!\mu'd\mu'\, n_0(\omega),
\end{align}
\end{subequations}
where we recall that $\omega=\gamma\omega_0\big(1+\beta\mu'\big)$.

Following similar steps for the single-mode particle, the condition of dynamical thermal equilibrium $\dot{Q}=0$ [see Eq.~(\ref{Pparticle})] becomes
\begin{align}\label{equilibrium22}
&0= \int_{-1}^1\frac{d\mu}{(1-\beta\mu)^2}\,\big[n_0(\omega)-n_1(\omega_0)\big].
\end{align}
Equivalently, as shown in Appendix~\ref{GisG}, this condition can be written as
\begin{align}\label{equilibrium222}
&0=\int_{-1}^1\!d\mu'\,\big[n_0(\omega)-n_1(\omega_0)\big],
\end{align}
which directly gives
\begin{align}\label{equilibrium2}
&n_1(\omega_0)=\frac{1}{2}\int_{-1}^1\,d\mu'\,n_0\big[\gamma\omega_0(1+\beta\mu')\big].
\end{align}
Note that $n_1(\omega_0)=n_0(\omega_0T_0/T_1)$, which shows explicitly that the temperature ratio $T_1/T_0$ is determined solely by the normalized velocity $\beta=v/c$ and the ratio of the resonance and thermal frequencies $\omega_0/\theta_0$. We plot the resulting $T_1/T_0$ in Fig.~\ref{Fig1}(d) as a function of these parameters. For moderate velocities, the particle becomes colder than the environment for resonance frequencies below $\sim0.4\,\theta_0$, and hotter for higher resonance frequencies. Substantial deviations of $T_1$ from $T_0$ occur mainly at relativistic velocities ($\beta\to1$), where $T_1/T_0$ can reach large values as $\omega_0/\theta_0$ increases [see Fig.~\ref{Fig1}(g)], presumably because strong Doppler shifts connect highly populated low-energy lab-frame photons ($\omega/\theta_0\to0$) to the rest-frame resonance at $\omega_0$.

\begin{figure}[t]
\centering\includegraphics[width=0.8\columnwidth]{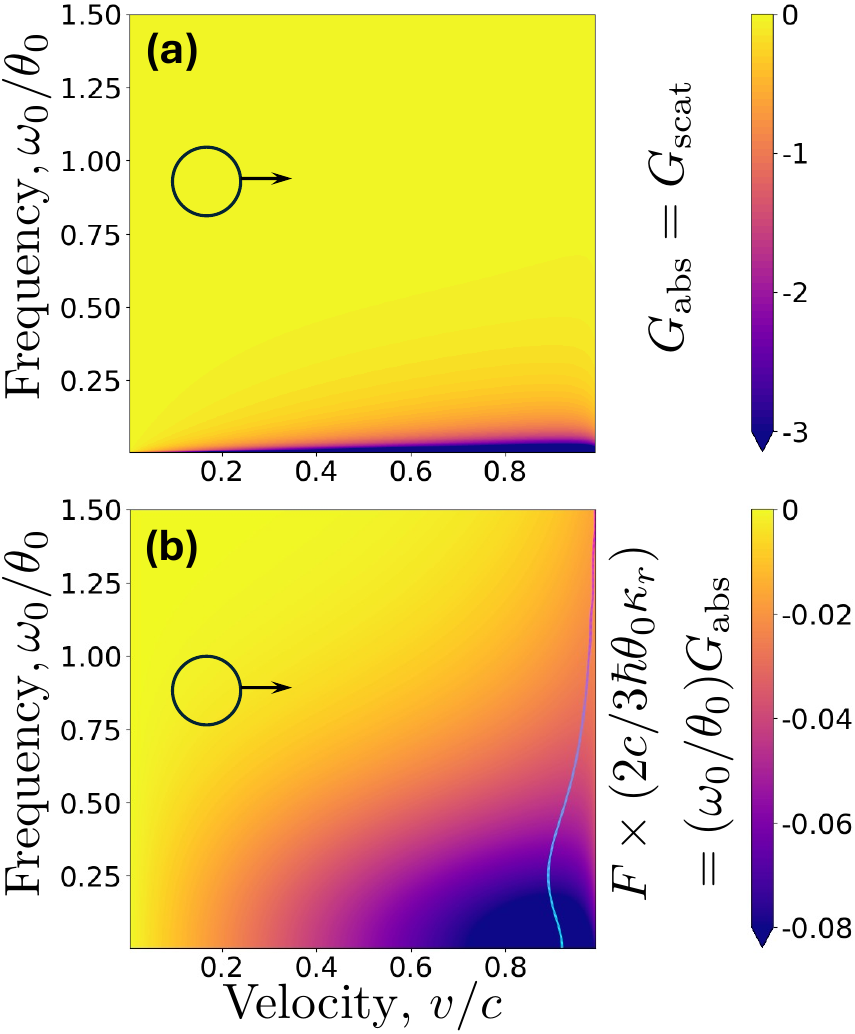}
\caption{Radiative thermal drag for a small isotropic particle moving in vacuum at thermal equilibrium. (a) We plot the dimensionless function $G_{\rm abs}=G_{\rm scat}$ defined in Eqs.~(\ref{GGGeq}), as a function of the normalized particle velocity and resonance frequency. (b) Normalized force under the conditions of (a). The solid curve in (b) marks the velocity that maximizes the force for each value of $\omega_0/\theta_0$.}\label{Fig2}
\end{figure}

Under thermal equilibrium, Eq.~(\ref{equilibrium222}) implies that the $\beta$ term vanishes in Eq.~(\ref{Gabseq}), while the $n_1(\omega_0)$ contribution also cancels because the integrand is odd in $\mu'$, therefore yielding $G_{\rm abs}=G_{\rm scat}$. Then, from Eqs.~(\ref{FFF}) and (\ref{Fparticlefinal}), the total force reduces to
\begin{align}\label{Ftotal}
&F\approx\frac{3\,\hbar\omega_0\,\kappa_r}{2c} G_{\rm abs}.
\end{align}
As expected, the force is proportional to $\kappa_r$, which encapsulates the coupling between the particle and the radiation field. Remarkably, the nonradiative contribution to the resonance linewidth, represented by the excess of $\kappa$ over $\kappa_r$, does not enter Eq.~(\ref{Ftotal}).

In Fig.~\ref{Fig2}, we show the dependence of $G_{\rm abs}$ [Eq.~(\ref{Gabseq})] and $F$ [Eq.~(\ref{Ftotal})] on the normalized velocity and resonance frequency assuming thermal equilibrium. As expected, the magnitudes of both $G_{\rm abs}$ and the resulting frictional force increase with velocity initially and reach their largest strength at $\beta\gtrsim0.9$, as shown by the solid curve in Fig.~\ref{Fig2}(b). The magnitude of $G_{\rm abs}$ also increases as $\omega_0/\theta_0$ decreases, reflecting the larger thermal occupation at lower resonance frequencies. In the force, this trend is partly offset by the additional factor of $\omega_0$ in Eq.~(\ref{Ftotal}).

\subsection{Extended homogeneous plate: out-of-plane motion}\label{vertical}

We again apply Eq.~(\ref{force}), but now the force is proportional to the plate area, which we take to be $L^2$ (i.e., the transverse area of the quantization box). It is then convenient to work with the frictional pressure $p=F/L^2$. For the plate, the absorption and scattering probabilities are $\pabs^\nu(\kb')=\mathcal{A}_\nu$ and $\pscat^\nu(\kb',\Omega'_s)=\mathcal{R}_\nu\delta(\tilde\Omega'-\Omega'_s)$, respectively, where $\mathcal{A}_\nu=1-|r_\nu(\omega',\mu')|^2-|t_\nu(\omega',\mu')|^2$ and $\mathcal{R}_\nu=|r_\nu(\omega',\mu')|^2$ are the absorbance and reflectance, which depend on the rest-frame quantities $\omega'$ and $\mu'$ through the reflection and transmission coefficients $r_\nu$ and $t_\nu$. Here, $\tilde\Omega'$ denotes the reflected direction associated with $\Omega'$, such that $\tilde\mu'=-\mu'$ and $\tilde\varphi'=\varphi'$.

Substituting these expressions into Eq.~(\ref{force}) and decomposing the pressure $p_\perp$ into absorption and scattering (or equivalently reflection) contributions as in Eq.~(\ref{FFF}), we obtain
\begin{align}\nonumber
p_\perp=p_{\rm abs}+p_{\rm scat}
\end{align}
with
\begin{align}\label{pplateperp}
\Bigg[\begin{split} &p_{\rm abs} \\ &p_{\rm scat} \end{split}\Bigg]
=\frac{\hbar}{4\pi^2c^3}\sum_\nu \int_0^\infty\!\!\omega^3d\omega\! \int_{-1}^1\! d\mu\, (1-\beta\mu)\,
\\\nonumber
\times
\Bigg[\begin{split}
&\mu\, \big[n_0(\omega)-n_1(\omega')\big]\, \big(1-|r_\nu(\omega',\mu')|^2-|t_\nu(\omega',\mu')|^2\big) \\
&(\mu-\tilde\mu\,\tilde\omega/\omega)\, n_0(\omega)\big[n_0(\tilde\omega)+1\big]\, |r_\nu(\omega',\mu')|^2
\end{split}\Bigg].
\end{align}
The scattering contribution in Eq.~(\ref{pplateperp}) involves the reflected frequency $\tilde\omega$ and polar direction $\tilde\mu$ in the lab frame. These quantities can be calculated as follows: denoting reflected quantities by a tilde, specular reflection in the rest frame $K'$ gives $\tilde\omega'=\omega'$ and $\tilde\mu'=-\mu'$; transforming back to the lab frame $K$ by replacing $\beta\to-\beta$ and exchanging primed and unprimed quantities in Eqs.~(\ref{transform}), we find $\tilde\omega=\gamma\omega'(1-\beta\mu')$ and $\tilde\mu=-(\mu'-\beta)/(1-\beta\mu')$; finally, substituting the expressions for $\omega'$ and $\mu'$ from Eqs.~(\ref{transform}), we obtain $\tilde\omega/\omega=\gamma^2(1-2\beta\mu+\beta^2)$ and $\tilde\mu=(2\beta-\mu-\beta^2\mu)/(1-2\beta\mu+\beta^2)$.

For illustration, we consider a thin plate described by a two-dimensional conductivity \cite{proc057}
\begin{align}\label{sigma}
\sigma(\omega')=\frac{e^2}{\hbar}\frac{\ii\,\omega_D}{\omega'-\omega_0+\ii\kappa/2},
\end{align}
corresponding to a single resonance frequency $\omega_0$ with oscillator strength $\omega_D$ and damping rate $\kappa$ (see Appendix~\ref{surfaceconductivity}). The Fresnel coefficients entering Eq.~(\ref{pplateperp}) are then given by \cite{GP16}
\begin{align}\nonumber
&t_s(\omega',\mu')=\frac{1}{1+2\pi\sigma(\omega')/c|\mu'|},
\\\nonumber
&t_p(\omega',\mu')=\frac{1}{1+2\pi\sigma(\omega')|\mu'|/c},
\end{align}
$r_s=t_s-1$, and $r_p=1-t_p$ for $\nu=\,$s and p polarizations. Note that the incidence direction enters these expressions through the absolute value $|\mu'|$. In the narrow-resonance limit $\kappa\ll\omega_0$, the absorbance and reflectance can be approximated as
\begin{align}\label{tr2D}
\begin{aligned}
&1\!-\!|r_\nu(\omega',\mu')|^2\!-\!|t_\nu(\omega',\mu')|^2
\\ &\quad\quad\quad\quad \to \pi\kappa\,\frac{\eta\xi_\nu}{1+\eta\xi_\nu}\, \delta(\omega'\!-\omega_0),
\\ &|r_\nu(\omega',\mu')|^2 \to \frac{\pi\kappa}{2}\,\frac{\eta^2\xi_\nu^2}{1+\eta\xi_\nu}\, \delta(\omega'\!-\omega_0),
\end{aligned}
\end{align}
where $\xi_s=1/|\mu'|$ and $\xi_p=|\mu'|$ for s and p polarization, respectively, and
\begin{align}\label{parameta}
\eta=\frac{4\pi\alpha\omega_D}{\kappa}
\end{align}
is a central parameter determining the relative strength of absorption and radiation contributions, proportional to the ratio of the transition strength $\omega_D$ to the damping rate $\kappa$ [see Eq.~(\ref{sigma})]. Here, $\alpha\approx1/137$ is the fine-structure constant. To derive Eqs.~(\ref{tr2D}), we have used transformations analogous to Eq.~(\ref{deltafunction}). Substituting these expressions into Eq.~(\ref{pplateperp}), carrying out the $\omega$ integral with the $\delta$ functions, using Eq.~(\ref{wpw}), and changing the integration variable from $\mu$ to $\mu'$ as shown in Appendix~\ref{GisG}, we obtain
\begin{align}\nonumber
\Bigg[\begin{split} &p_{\rm abs} \\ &p_{\rm scat} \end{split}\Bigg]
\approx\frac{\alpha\hbar\omega_D\omega_0^3}{c^3}\sum_\nu \int_{-1}^1\! d\mu'\,(1+\eta\xi_\nu)^{-1}\,\xi_\nu
\\\label{pplateperp2}
\times
\Bigg[\begin{split}
&(\beta+\mu')\, \big[n_0(\omega)-n_1(\omega_0)\big] \\
&\eta\,\mu'\xi_\nu\, n_0(\omega)\big[n_0(\tilde\omega)+1\big]
\end{split}\Bigg]
\end{align}
for the absorption and scattering contributions to the frictional pressure. Here, we must use $\omega=\gamma\omega_0(1+\beta\mu')$ and $\tilde\omega=\gamma\omega_0\,(1-\beta\mu')$, as obtained from Eqs.~(\ref{wpw}) and (\ref{wsw}) with $\tilde\mu'=-\mu'$.

Applying the same procedure to the thermal equilibrium condition in Eq.~(\ref{equilibrium}), we find
\begin{align}\label{equilibriumperp}
0=\sum_\nu\int_{-1}^1 d\mu'\,(1+\eta\xi_\nu)^{-1}\,\xi_\nu\, \big[n_0(\omega)-n_1(\omega_0)\big].
\end{align}
In Fig.~\ref{Fig1}(e), we plot the resulting temperature ratio $T_1/T_0$ for $\eta=0.4$. Compared with Fig.~\ref{Fig1}(d), the isothermal contours are shifted slightly upward, presumably because of the different angular distribution of photon scattering. In the $\beta\lesssim1$ regime, high $T_1/T_0$ ratios are observed [Fig.~\ref{Fig1}(g)], similar to those found for the particle.

\begin{figure*}[t]
\includegraphics[width=1.0\textwidth]{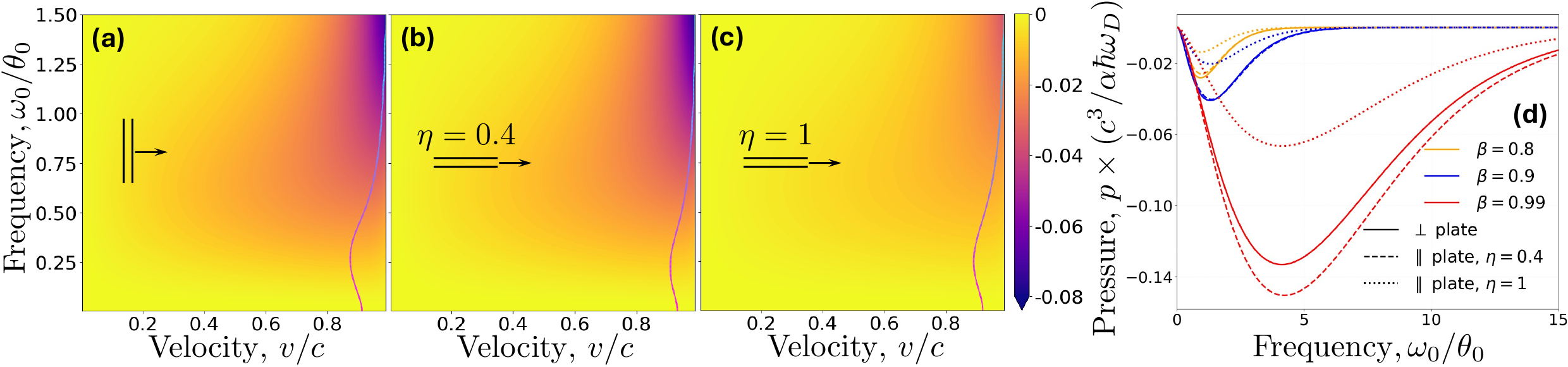}
\caption{Radiative thermal drag for extended homogeneous plates moving in vacuum at thermal equilibrium as functions of the normalized plate velocity and resonance frequency. Solid curves mark the velocity of maximum pressure for each value of $\omega_0/\theta_0$. In panel (a), we plot the normalized pressure $p_\perp$ for the out-of-plane moving plate [Eq.~(\ref{eqPperp})], which is independent of $\eta$. In panels (b) and (c), we plot the normalized pressure $p_\parallel$ for in-plane moving plates characterized by $\eta=0.4$ and $\eta=1$, respectively. Panel (d) shows the normalized pressure as a function of resonance frequency under the conditions of panels (a)--(c) for different values of $\beta$ (see legend).}
\label{Fig3}
\end{figure*}

Under thermal equilibrium, Eq.~(\ref{equilibriumperp}) implies that the term proportional to $\beta$ in $p_{\rm abs}$ vanishes [see Eq.~(\ref{pplateperp2})], as does the contribution proportional to $n_1(\omega_0)$, in analogy with the particle case. Combining the absorption and scattering contributions, and using the odd parity of the $n_0(\omega)n_0(\tilde\omega)$ term as a function of $\mu'$ in the scattering integral, we find
\begin{align}\nonumber
p_\perp &\approx \frac{\alpha\hbar\omega_D\omega_0^3}{c^3}\sum_\nu\int_{-1}^1\! d\mu'\,\xi_\nu\,\mu'\,n_0(\omega) \nonumber \\
&= \frac{\alpha\hbar\omega_D\omega_0^3}{c^3}\int_{-1}^1\! d\mu'\,n_0(\omega)\,{\rm sgn}(\mu')\Big(1+\mu'^2\Big). \label{eqPperp}
\end{align}
The total frictional pressure acting on the normally moving plate is therefore independent of $\eta$ [Fig.~\ref{Fig3}(a)]. This result originates from the complementary momentum transfer produced by absorption and reflection. For normal motion, both processes exchange momentum along the direction of the plate velocity. Changing $\eta$ modifies the partition of the resonant response between these two channels, but the reduction of the absorption contribution is compensated by the corresponding increase of the reflection contribution. Under thermal equilibrium, their sum therefore becomes independent of this partition and is determined only by the overall strength of the resonant optical response. This cancellation requires both absorption and reflection and, as shown below, does not occur for parallel motion, where reflection transfers no momentum along the direction of motion. For a fixed resonance frequency $\omega_0$, the magnitude of the drag pressure increases with velocity and reaches its largest absolute value at high velocities. In contrast to the radiative force acting on a particle, at fixed velocity, $|p_\perp|$ displays a maximum at a finite value of $\omega_0/\theta_0$. This optimal resonance frequency shifts upward as $\beta\to1$, accompanied by an increase in the pressure magnitude [Fig.~\ref{Fig3}(d)]. This effect correlates with the relativistic increase in the equilibrium plate temperature [Fig.~\ref{Fig1}(g)] and benefits from the higher population of lower-frequency thermal photons in the vacuum bath, which are Doppler-shifted into resonance with the material response.

\subsection{Extended homogeneous plate: in-plane motion}\label{horizontal}

The problem of radiative friction simplifies when the plate moves along an in-plane direction, taken here as the $z$ axis in the coordinate system of Fig.~\ref{Fig1}(a), with the surface normal along the $x$ axis [see Fig.~\ref{Fig1}(c)]. In this geometry, specular reflection reverses the normal component $k'_x$ of the photon wave vector while leaving its component $k'_z$ along the direction of motion unchanged. Consequently, the reflected frequency is unchanged in the lab frame and reflection does not contribute to the frictional pressure, since $(\kb-\kb_s)\cdot\zz=0$ in Eq.~(\ref{force}), irrespective of polarization.

For evaluating the absorption contribution, it is convenient to use the TE and TM polarization basis defined with respect to the plate normal, in which the optical response of the isotropic plate is diagonal. In the present geometry, this basis generally differs from the p and s basis defined in Sec.~\ref{kinematics} with respect to the velocity direction. However, because the incident thermal radiation is unpolarized and Eq.~(\ref{force}) involves a sum over the two incident polarization states, this sum is invariant under a change of polarization basis. We can therefore evaluate it in the TE/TM basis of the plate. Defining the corresponding absorption probabilities as $\mathcal{A}_\lambda(\omega',\mu'_x)
=1-|r_\lambda(\omega',\mu'_x)|^2-|t_\lambda(\omega',\mu'_x)|^2$ with $\lambda=$TE, TM, where $\mu'_x=\sqrt{1-\mu'^2}\cos\varphi$ is the cosine of the incidence angle with respect to the surface normal, using Eq.~(\ref{force}) to calculate the force, and dividing by the plate area $L^2$, we obtain the frictional pressure
\begin{align}\label{pressurepara0}
p_\parallel&=\frac{\hbar}{(2\pi c)^3}\int_0^{\infty}\omega^3d\omega\,
\int_{-1}^1 \mu\,d\mu\, (1-\beta\mu)
\\\nonumber
&\times \big[n_0(\omega) - n_1(\omega')\big]
\sum_{\lambda={\rm TE,TM}}\int_0^{2\pi}\!\!\!d\varphi\, \mathcal{A}_\lambda(\omega',\mu'_x).
\end{align}
Assuming the same single-resonance thin-plate response as in Sec.~\ref{vertical}, we use the Fresnel coefficients given there with s and p now corresponding, respectively, to the TE and TM polarizations defined with respect to the plate normal. Equation~(\ref{pressurepara0}) then reduces to
\begin{align}\label{pressurepara}
p_\parallel\approx\frac{2\alpha\hbar\omega_D\omega_0^3}{\pi c^3}
&\int_{-1}^1 d\mu'\, (\beta+\mu')\, \big[n_0(\omega)-n_1(\omega_0)\big]
\\\nonumber
&\times \int_0^{\pi/2}\!\!\!d\varphi\,\bigg(\frac{1}{\mu'_x+\eta}+\frac{\mu'_x}{1+\eta\mu'_x}\bigg),
\end{align}
where $\omega=\gamma\omega_0(1+\beta\mu)$ and $\eta$ is defined in Eq.~(\ref{parameta}).

Applying the same procedure to Eq.~(\ref{equilibrium}), the condition for thermal equilibrium takes a form analogous to Eq.~(\ref{equilibriumperp}), we obtain
\begin{align}\label{equilibriumpara}
0=&\int_{-1}^1 d\mu'\, \big[n_0(\omega)-n_1(\omega_0)\big]
\\\nonumber
&\times \int_0^{\pi/2}\!\!\!d\varphi\,\bigg(\frac{1}{\mu'_x+\eta}+\frac{\mu'_x}{1+\eta\mu'_x}\bigg).
\end{align}
In Fig.~\ref{Fig1}(f), we show the resulting temperature ratio $T_1/T_0$ for $\eta=0.4$. Relative to the particle and to the plate moving out of plane, the isothermal contours are shifted slightly downward as a function of $\omega_0/\theta_0$, while the $\beta\lesssim1$ behavior remains essentially unchanged [Fig.~\ref{Fig1}(g)].

As in the out-of-plane geometry, Eq.~(\ref{equilibriumpara}) implies that the term proportional to $\beta$ in $p_\parallel$ vanishes under thermal equilibrium [see Eq.~(\ref{pressurepara})], as does the contribution proportional to $n_1(\omega_0)$. Unlike in the out-of-plane case, however, the remaining terms in $p_\parallel$ retain a strong dependence on $\eta$.

This behavior reflects the absence of a contribution from elastic reflection to the frictional pressure in the parallel geometry: specular reflection leaves the photon momentum component along the direction of motion unchanged, so that only absorption and emission contribute to the drag. Consequently, the pressure retains a strong dependence on $\eta$. Larger values of $\eta$, corresponding to stronger relative radiation coupling, lead to weaker thermal drag [Figs.~\ref{Fig3}(b,c)]. Its relativistic evolution follows the same qualitative trend as in the small-particle and perpendicular-plate geometries, with an upward shift of the optimal resonance frequency and an increase in the pressure magnitude as $\beta\to1$ [Fig.~\ref{Fig3}(d)].

\subsection{Low-velocity friction coefficients}
\label{lowv}

We consider thermal equilibrium and focus on a small particle and thin plates with a single-frequency response. In the $\beta\ll1$ limit, the force $F$ on the particle and the pressures $p_\perp$ and $p_\parallel$ on the plates are expected to scale linearly with $\beta$. Under thermal equilibrium conditions, the corresponding friction coefficients are then given by the low-$\beta$ expansion of Eq.~(\ref{Ftotal}) for the particle and Eqs.~(\ref{pplateperp2}) and (\ref{pressurepara}) for the plates moving normal or parallel to their surfaces, respectively.

The three equilibrium conditions [Eqs.~(\ref{equilibrium222}), (\ref{equilibriumperp}), and (\ref{equilibriumpara})] share the structure $0=\int_{-1}^{1}\!d\mu'\,w(\mu')\,[n_0(\omega)-n_1(\omega_0)]$, with an angular weight $w(\mu')$ that is even in $\mu'$. To first order in $\beta$, one has $\gamma^{-4}\approx1$ and $\omega\approx\omega_0(1+\beta\mu')$, and therefore, $n_0(\omega)\approx n_0(\omega_0)+\beta\mu'\,\omega_0 n_0'(\omega_0)$, where $n_0'(\omega_0)=-(\hbar/\kB T_0)\,n_0(\omega_0)[n_0(\omega_0)+1]$ is the derivative with respect to frequency. Inserting these expressions in the equilibrium conditions, the term linear in $\beta$ is odd in $\mu'$ and integrates to zero, so to first order, $n_1(\omega_0)\approx n_0(\omega_0)$ and $T_1\approx T_0$. Hence,
\begin{align}\label{n0n1plates}
n_0(\omega)-n_1(\omega_0)\approx\beta\mu'\,\omega_0\,n_0'(\omega_0)
\end{align}
inside the force and pressure integrals of Eqs.~(\ref{Gabseq}), (\ref{pplateperp2}), and (\ref{pressurepara}).

\textit{Small particle.}---Inserting Eq.~(\ref{n0n1plates}) into Eq.~(\ref{Gabseq}), we obtain $G_{\rm abs}\approx(2/3)\beta\,\omega_0\,n_0'(\omega_0)$, and from Eq.~(\ref{Ftotal}),
\begin{align}\label{Fsmallv}
&\frac{F}{\beta}\approx-\frac{(\hbar\omega_0)^2\,\kappa_r}{c\,\kB T_0}\, n_0(\omega_0)\big[n_0(\omega_0)+1\big],
\end{align}
where we have expressed $n_0'(\omega_0)$ in terms of $n_0(\omega_0)$. This result is consistent with the universal macroscopic thermal drag obtained by Mkrtchian \textit{et al.} \cite{MPP03}. By applying the identity $1/\sinh^2(\hbar\omega/2\kB T_0) = 4n_0(\omega)[n_0(\omega)+1]$ and mapping their MKS susceptibility to a single-particle polarizability in Gaussian units via $V\chi_e''(\omega) = 4\pi {\rm Im}\{\alpha(\omega)\}$, their drag equation reduces exactly to Eq.~(\ref{Fsmallv}) in the narrow-linewidth limit ${\rm Im}\{\alpha(\omega)\} \to (3\pi c^3 \kappa_r/4\omega_0^3)\delta(\omega-\omega_0)$. In the present formulation, the absorption and scattering contributions are explicitly separated, allowing us to directly examine the pure-scattering limit $\kappa\to\kappa_r$. By explicitly incorporating the momentum transfer from individual scattering events, our model retains the scattering contribution to the frictional drag in this limit. The origin of this finite force can be understood from the Doppler shift associated with photon scattering. Although scattering is elastic in the particle rest frame, it is generally inelastic in the lab frame because the transformation back to this frame depends on the scattering direction [see Eq.~(\ref{wsw})]. The corresponding change in photon energy and momentum is balanced by the translational motion of the particle. In addition, photons arriving from different directions experience different Doppler shifts, resulting in an imbalance between their contributions to the momentum transfer. Consequently, a finite stopping force remains even in the absence of nonradiative losses.

\textit{Plate in out-of-plane motion.}---Inserting Eq.~(\ref{n0n1plates}) into Eq.~(\ref{pplateperp2}) and keeping the leading term in $\beta$, the absorption and scattering pressures become
\begin{align}\nonumber
\frac{1}{\beta}\Bigg[\begin{matrix} p_{\rm abs} \\ \\ p_{\rm scat} \end{matrix}\Bigg]
\approx&
\frac{\alpha\hbar\omega_D\omega_0^4}{c^3}\, n_0'(\omega_0)
\\\nonumber
&\times\sum_\nu\!\int_{-1}^{1}\!\mu'^2d\mu'\,\frac{\xi_\nu}{1+\eta\xi_\nu}\;
\Bigg[\begin{matrix} 1 \\ \\ \eta\xi_\nu \end{matrix}\Bigg],
\end{align}
where in the scattering term we have used $\tilde\omega\approx\omega_0(1-\beta\mu')$, so that $n_0(\omega)[n_0(\tilde\omega)+1]\approx n_0(\omega_0)[n_0(\omega_0)+1]+\beta\mu'\omega_0 n_0'(\omega_0)$ and only the part linear in $\beta$ survives the (otherwise odd) $\mu'$ integral. Adding the two contributions, the dependence on $\eta$ cancels exactly, and we obtain
\begin{align}\label{pperplowv}
\frac{p_\perp}{\beta}\approx-\frac{3\alpha\hbar^2\omega_D\omega_0^4}{2c^3\kB T_0}\, n_0(\omega_0)\big[n_0(\omega_0)+1\big]
\end{align}

Remarkably, as in the particle case, the result is independent of nonradiative losses (i.e., independent of $\eta$). This cancellation can be seen directly from the absorption and scattering terms above: their relative weights $1$ and $\eta\xi_\nu$ add to the factor $1+\eta\xi_\nu$, which cancels the same factor in the denominator. The total pressure is therefore independent of how the resonant response is distributed between absorption and reflection, so it depends only on the oscillator strength $\omega_D$. It is useful to contrast Eq.~(\ref{pperplowv}) with previous calculations of thermal drag on perfectly reflecting mirrors moving normal to their surfaces \cite{MMF02}. For idealized broadband mirrors in the low-velocity limit, the Doppler shift of reflected photons across the full thermal spectrum produces a macroscopic frictional pressure proportional to $T_0^4$. In contrast, for the thin plate considered here, whose response is dominated by a narrow resonance at $\omega_0$, the low-velocity momentum transfer is controlled by the thermal factor $n_0(\omega_0)[n_0(\omega_0)+1]$. This comparison underscores the central role of the material optical response in vacuum friction.

\textit{Plate in parallel motion.}---Inserting Eq.~(\ref{n0n1plates}) into Eq.~(\ref{pressurepara}), we obtain
\begin{align}\nonumber
\frac{p_\parallel}{\beta}\approx-\,\frac{2\alpha\hbar^2\omega_D\omega_0^4}{\pi c^3\kB T_0}\,
\mathcal{I}_\parallel(\eta)\,n_0(\omega_0)\big[n_0(\omega_0)+1\big],
\end{align}
where we have introduced the dimensionless factor
\begin{align}\nonumber
\mathcal{I}_\parallel(\eta)=\int_{-1}^{1}\!d\mu'\,\mu'^2
\int_0^{\pi/2}\!d\varphi
\left[\frac{1}{\mu'_x+\eta}+\frac{\mu'_x}{1+\eta\mu'_x}\right].
\end{align}
In contrast to the normal-incidence scenario, the pressure retains a dependence on $\eta$ because the elastic reflection does not contribute to the frictional pressure in the parallel geometry. The $\mathcal{I}_\parallel(\eta)$ factor decreases monotonically with $\eta$, takes the value $\mathcal{I}_\parallel(1)=\pi/3$, and has limiting forms $\mathcal{I}_\parallel(\eta)\;\xrightarrow{\;\eta\gg1\;}\;2\pi/3\eta$ and $\mathcal{I}_\parallel(\eta)\;\xrightarrow{\;\eta\ll1\;}\;-(\pi/2)\,\ln\eta$. Interestingly, $p_\perp$ and $p_\parallel$ coincide only when $\mathcal{I}_\parallel(\eta)=3\pi/4$, which is obtained for $\eta\approx0.25$, while the in-plane drag exceeds the out-of-plane one for smaller $\eta$.

\section{Conclusions}\label{conclusions}

In this work, we have developed a kinetic momentum-transfer theory of the radiative drag experienced by objects moving through a thermal photon bath. By evaluating the momentum exchanged in individual photon absorption, emission, and scattering events, this approach provides an intuitive and self-consistent description of vacuum friction. It naturally incorporates relativistic Doppler effects and enables the material optical response to be treated in the appropriate rest frame while the resulting force or pressure is evaluated in the lab frame. We have applied this formalism to three representative geometries: an isotropic dipolar particle, a thin plate moving normal to its surface, and a thin plate moving along an in-plane direction. In all cases, the material response is assumed to be dominated by a single resonance at frequency $\omega_0$. Under dynamical equilibrium with the surrounding photon bath of thermal frequency $\theta_0=2\pi\kB T_0/\hbar$, the temperature of the moving object can lie either above or below the vacuum temperature, depending on the velocity and on $\omega_0$. In the three geometries considered here, the object-to-vacuum temperature ratio follows a similar qualitative trend, such that it is close to unity near $\omega_0/\theta_0\sim0.5$ at moderate velocities and increases substantially in the relativistic regime as $\beta\to1$.

As a function of velocity, the frictional drag (i.e., a force for the particle and a pressure for the plates) reaches its largest magnitude at relativistic velocities, typically for $v\gtrsim0.9\,c$, in all three geometries. Its dependence on resonance frequency, however, differs markedly between small particles and thin plates. For the particle, the largest drag values occur at low resonance frequencies, reflecting the dominant role of the Bose--Einstein occupation factor $n_0(\omega_0)$ at the bath temperature $T_0$. In contrast, for the plates, the drag magnitude is maximized at a finite value of $\omega_0/\theta_0$. At moderate velocities, this optimum value is consistent with the maximum of the factor $\omega_0^3\,n_0(\omega_0)$, analogous to the spectral weighting in blackbody emission. In the relativistic regime, the maximum becomes more pronounced and shifts to larger values of $\omega_0/\theta_0$ as $\beta\to1$. As in the relativistic enhancement of the equilibrium object temperature, this behavior originates from Doppler shifts that couple the material resonance to the highly populated low-energy sector of the thermal photon bath. For a plate moving normal to its surface, the resulting frictional pressure is independent of the relative absorption strength, provided that both absorption and scattering channels are retained. This is in contrast to the parallel-motion geometry, where elastic photon scattering does not contribute to the drag and the pressure therefore retains a strong dependence on the absorption strength.

The present approach applies to systems for which the interaction with thermal radiation can be described in terms of the absorption and elastic-scattering probabilities considered here. Extension to configurations involving, for example, near-field coupling between different bodies would require a corresponding generalization of the momentum-transfer description.
\\

\acknowledgements
This work has been supported in part by the European Research Council (101141220-QUEFES), the Spanish MICINN (PID2024-157421NB-I00 and Severo Ochoa CEX2024-001490-S), and the CERCA program. G.C. acknowledges support from the "la Caixa" Foundation (ID 100010434) through fellowship LCF/BQ/DFI26/13001086.
\\

\appendix

\section{Proof of Eqs.~(\ref{GGGeq}) and (\ref{equilibrium222})}\label{GisG}
\renewcommand{\theequation}{A\arabic{equation}}

We start by changing the integration variable from $\mu$ to $\mu'$ in Eqs.~(\ref{GGG}). Specifically, we write $\mu=(\mu'+\beta)/(1+\beta\mu')$, which is the counterpart of Eq.~(\ref{mupmu}) for a $-v$ boost. Then, Eq.~(\ref{Gabs}) readily becomes Eq.~(\ref{Gabseq}). Further applying the analogous transformation $\mu_s\to\mu'_s$ to Eq.~(\ref{Gscat}), we obtain
\begin{align}\nonumber
&G_{\rm scat}=\frac{3}{16} \int_{-1}^1\!d\mu'\int_{-1}^1\!d\mu'_s\,(\mu'-\mu'_s)
\\\nonumber
&\quad\quad\quad\times
\big(3-\mu'^2-\mu'^2_s+3\mu'^2\mu'^2_s\big)\, n_0(\omega)\, \big[n_0(\omega_s)+1\big].
\end{align}
Applying the transformation in Eq.~(\ref{wpw}) to both $\omega$ and $\omega_s$, and using the fact that the incident and scattered frequencies are equal in the rest frame, we find that the term in the integrand proportional to $n_0(\omega)\,n_0(\omega_s)$ is antisymmetric under the exchange $\mu'\leftrightarrow\mu'_s$. Therefore, its contribution vanishes. In the remaining term, the $\mu'_s$ integral can be carried out directly, yielding Eq.~(\ref{Gscateq}). Finally, applying the same change of variables to Eq.~(\ref{equilibrium22}) directly produces the equilibrium condition in Eq.~(\ref{equilibrium222}).

\section{Surface conductivity of a thin single-resonance plate}\label{surfaceconductivity}
\renewcommand{\theequation}{B\arabic{equation}}

We consider a material characterized by a permittivity $\epsilon(\omega)=\epsilon_\infty-(\epsilon_0-\epsilon_\infty)\omega_0^2/[\omega(\omega+\ii\kappa)-\omega_0^2]$. This Lorentz-type form is commonly used for phonon-polaritonic materials, for which $\epsilon_0-\epsilon_\infty\lesssim10$ \cite{AM1976}. It may also describe a metamaterial engineered to have a sufficiently low resonance frequency $\omega_0$, comparable to the thermal frequency in the present context.

We describe a thin plate of thickness $d$ in terms of an effective surface conductivity $\sigma(\omega)$, which is obtained by comparing the bulk permittivity with $\epsilon(\omega)=\epsilon_\infty+4\pi\ii\sigma(\omega)/\omega d$ and neglecting the nonresonant $\epsilon_\infty$ contribution. In the narrow-linewidth limit $\kappa\ll\omega_0$, this procedure yields Eq.~(\ref{sigma}) with an oscillator strength $\omega_D=f\omega_0^2d/c$, where $f=(\epsilon_0-\epsilon_\infty)/8\pi\alpha\sim0-55$. From Eq.~(\ref{parameta}), we also find $\eta=(1/2)(\epsilon_0-\epsilon_\infty)(\omega_0 d/c)(\omega_0/\kappa)$, where $\omega_0 d/c\ll1$ in the thin-film limit, but $\omega_0/\kappa\gg1$ in a low-loss material, so we can encounter a large range of $\eta$ values.


%

\end{document}